\pdfoutput=1

\documentclass[11pt]{article}

\usepackage[]{acl}

\usepackage{times}
\usepackage{latexsym}

\usepackage{amsmath,graphicx}
\usepackage[T1]{fontenc}

\usepackage[utf8]{inputenc}

\usepackage{microtype}

\usepackage{arydshln}
\usepackage{multirow}
\usepackage{placeins}
\usepackage{booktabs}

\usepackage[T1]{fontenc}

\usepackage[utf8]{inputenc}

\usepackage{microtype}

\usepackage{inconsolata}

\makeatletter
\newcommand{\printfnsymbol}[1]{%
  \textsuperscript{\@fnsymbol{#1}}%
}
\makeatother

\usepackage{multicol}

\usepackage{listings}
\usepackage{xcolor}

\definecolor{amber}{rgb}{1.0, 0.75, 0.0}

\usepackage[normalem]{ulem}
\useunder{\uline}{\ul}{}

\definecolor{background}{HTML}{EEEEEE}

\usepackage{xcolor}

\lstdefinelanguage{json}{
    basicstyle=\normalfont\ttfamily,
    numbers=left,
    numberstyle=\scriptsize,
    stepnumber=1,
    numbersep=8pt,
    showstringspaces=false,
    breaklines=true,
    frame=lines,
    moredelim=**[is][\color{orange}]{@}{@},
    moredelim=**[is][\color{amber}]{!}{!},
    backgroundcolor=\color{background},
    literate=
     *{0}{{{\color{numb}0}}}{1}
      {1}{{{\color{numb}1}}}{1}
      {2}{{{\color{numb}2}}}{1}
      {3}{{{\color{numb}3}}}{1}
      {4}{{{\color{numb}4}}}{1}
      {5}{{{\color{numb}5}}}{1}
      {6}{{{\color{numb}6}}}{1}
      {7}{{{\color{numb}7}}}{1}
      {8}{{{\color{numb}8}}}{1}
      {9}{{{\color{numb}9}}}{1}
      {:}{{{\color{punct}{:}}}}{1}
      {,}{{{\color{punct}{,}}}}{1}
      {\{}{{{\color{delim}{\{}}}}{1}
      {\}}{{{\color{delim}{\}}}}}{1}
      {[}{{{\color{delim}{[}}}}{1}
      {]}{{{\color{delim}{]}}}}{1},
}

\title{1SPU: 1-step Speech Processing Unit}

\author{Karan Singla,
        Shahab Jalavand,
        Andrej Ljolje,
        Antonio Moreno Daniel,\\
        {\bf Srinivas Bangalore},
        {\bf Yeon-Jun Kim},
        {\bf Ben Stern}\\
        karan@whissle.ai, sjalalvand@interactions.com}

\begin{document}


\maketitle
\begin{abstract}
Recent studies have made some progress in refining end-to-end (E2E) speech recognition encoders by applying Connectionist Temporal Classification (CTC) loss to enhance named entity recognition within transcriptions. However, these methods have been constrained by their exclusive use of the ASCII character set, allowing only a limited array of semantic labels. We propose 1SPU, a 1-step Speech Processing Unit which can recognize speech events (e.g: speaker change) or an NL event (Intent, Emotion) while also transcribing vocal content. It extends the E2E automatic speech recognition (ASR) system's vocabulary by adding a set of unused placeholder symbols, conceptually akin to the <pad> tokens used in sequence modeling. These placeholders are then assigned to represent semantic events (in form of tags) and are integrated into the transcription process as distinct tokens. 

We demonstrate notable improvements on the SLUE benchmark and yields results that are on par with those for the SLURP dataset. Additionally, we provide a visual analysis of the system's proficiency in accurately pinpointing meaningful tokens over time, illustrating the enhancement in transcription quality through the utilization of supplementary semantic tags.

\end{abstract}

\section{Introduction}

This paper extends the capabilities of End-to-End Automatic Speech Recognition (E2E ASR) systems from traditional transcription tasks to more nuanced speech understanding processes \cite{chan2015listen, bahdanau2016end}. Presently, Deep Neural Networks (DNNs) are proficient in applying greedy or beam search decoding to predict vocabulary labels from audio inputs over time \cite{watanabe2017hybrid, he2019streaming}. These networks typically produce a sub-word text token for each segment of a continuous speech stream, focusing on transcription fidelity.

However, the transcription process has evolved with recent efforts to embed additional metadata, such as named entity boundaries and utterance classification \cite{serdyuk2018towards, haghani2018audio}. Multitask learning frameworks utilizing separate loss functions for transcription and classification have emerged \cite{wang2023end, ghannay2018end}, and there has been success in identifying sequences of intents using non-autoregressive models within a speech stream \cite{potdar2021streaming}. Although recognizing a limited set of named entities using CTC loss by marking them with specific begin and end tokens in transcriptions has been demonstrated \cite{ghannay2018end}. Methods like these are often use special ASCII characters for events, constrained by the size ASCII character set. Morever these special characters are originally defined to be useful for meaningful language transcription. This also happens a as ASR encoders are generally considered as intermediaries between speech and natural language (NL) models.

In response, our study proposes a single-step, integrated speech processing unit (1SPU) using E2E speech encoders. 1SPU  not only transcribes spoken content but also annotates it with tags indicating various NLP and speech events, including intent and entity recognition as well as speaker changes. By incorporating $user\_defined\_symbols$ into the ASR vocabulary \cite{kudo2018sentencepiece}, akin to text-based entity tagging tokens, we enhance the encoder's lexicon. These symbols are introduced during the fine-tuning stage and are only used if present in the fine-tuning data, thereby maintaining computational efficiency. We fine-tune a pre-trained Conformer ASR model to perform transcription while also producing intent labels and entity tags, all under the CTC loss function. The model outputs transcriptions using standard greedy decoding and can function in both offline and streaming modes.

Our experimentation focuses on mono-channel speech and yields promising results on the SLUE \cite{wang2021voxpopuli} and SLURP \cite{bastianelli2020slurp} benchmarks. We posit that incorporating a Language Model (LM) could significantly elevate these results. Furthermore, we present visual evidence that utilizing context from previous utterances can enhance transcription and understanding in dual-channel human-human interactions, particularly in marking speaker changes. This ongoing work has the potential to improve prompting strategies and facilitate the development of multi-modal and multitasking E2E ASR applications in streaming contexts.

The contributions of this paper are summarized as follows:

\begin{itemize}
\item We propose 1SPU, which incorporates the central idea of $user\_defined\_symbols$ into the ASR vocabulary to facilitate the tagging of speech and NL events within continuous speech transcriptions during fine-tuning.
\item We demonstrate that our method not only improves performance on the SLUE benchmark but also shows promise for the SLURP dataset.
\item We offer visual proof that the inclusion of context from prior events can improve speech transcription and understanding in scenarios requiring low-latency responses.
\end{itemize}

\section{Related Work}

A common practice is to transform the normalized token sequences produced by automatic speech recognition (ASR) systems into a \emph{written form} that is more amenable for downstream dialog system components to process \cite{pusateri2017mostly}. This reformatted written output then serves as the basis for extracting \emph{structured information}, such as intents and slot values, which are crucial for the continuation of a dialog \cite{radfar2020end}. Of late, there has been an emerging trend towards utilizing neural encoders that are optimized directly on speech inputs, a methodology often referred to as \emph{end-to-end spoken language understanding (E2E SLU)} \cite{serdyuk2018towards,haghani2018audio}.

{\bf Text-based Sequence Tagging: } The domain of natural language processing has seen extensive research in employing sequence-to-sequence (seq-2-seq) models to parse and extract structured information from textual data. This encompasses advancements in identifying the positions and classifications of entities within a text \cite{ratinov2009design}, generating structured summaries of content \cite{gorinski2019named}, and constructing synctactic parse trees through timely tag predictions during the generation of language \cite{cabot2021rebel}. Despite their efficacy, these approaches, which are primarily designed for monologic text, do not adequately address the dynamic nature of spoken dialogue.

{\bf E2E ASR:} Modern Automatic Speech Recognition (ASR) systems excel at converting spoken language into a sequence of lexical tokens, effectively transcribing spoken words. This transcription process typically involves refining pre-trained self-supervised neural networks or initiating training from the ground up with non-autoregressive Connectionist Temporal Classification (CTC) loss  or a sequential loss \cite{graves2012sequence} to recognize speech tokens \cite{chan2015listen, chorowski2015attention}. Transcriptions are obtained either by greedy decoding or beam-search decoding using a language model.

{\bf E2E SLU:} With the emergence of end-to-end ASR \cite{chorowski2015attention, chan2016listen} and the successful pretraining of speech encoders, methods for SLU directly from the speech signal have recently shown comparable performance to the conventional approach of cascading ASR and text-based components in tasks such as named entity recognition (NER), translation, dialogue act prediction (DAP) \cite{vila2018end, dang2020end}, as well as inference tasks like emotion, intent or behavior understanding \cite{fayek2015towards, price2020improved,singla2020towards}. Recent studies have attempted to identify the beginning and end of named entities in transcriptions by adding unique markers to the text  \cite{ghannay2018end, tomashenko2019investigating}. However, the performance of these methods is often limited by the small selection of special characters available in the ASCII character set and fear of using special symbols which may conflict with making multi-lingual ASR systems or other desired symbols.

\section{1SPU: 1-step Speech Processing Unit}

\begin{figure}[htbp]
\centering
    \includegraphics[width=53.3mm,height=65mm]{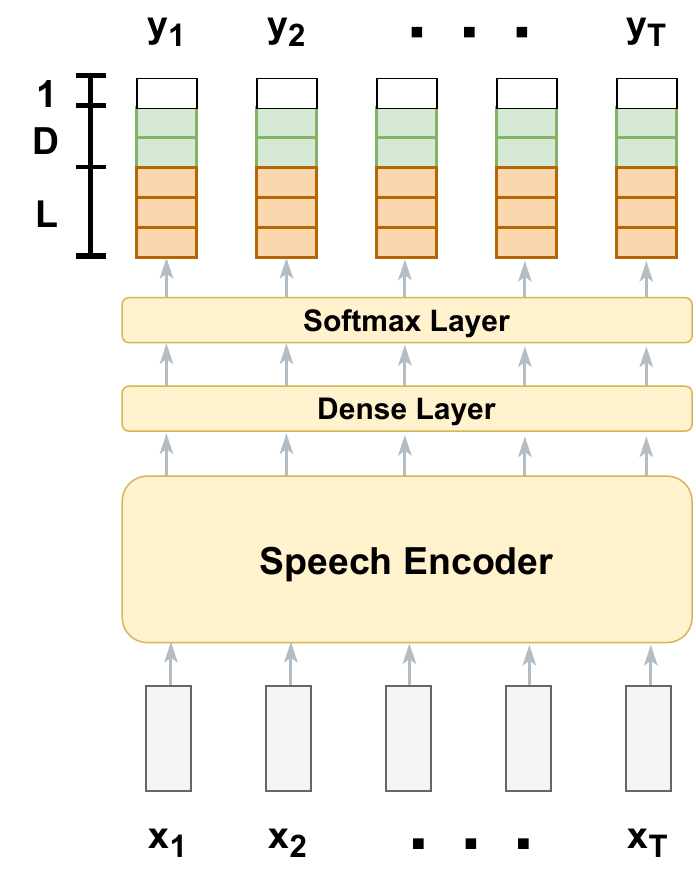}

\caption{Overview of 1SPU. $L$ refers to transcription tokens, $D$ refers to un-used tokens learnt during fine-tuning to mark events. $1$ is the \emph{blank} token.}
\end{figure}

For 1-step Speech Processing Unit (1SPU) we leverage a commercially available End-to-End Automatic Speech Recognition (E2E ASR) system, initializing it with a tokenizer that includes $D$ user-defined dummy tokens. These labels are akin to unique tokens, much like a `<pad>` token, and are not utilized during the ASR's initial training phase. Thus, the incorporation of these inactive tokens does not impact the duration of training or inference as they can be masked during the E2E ASR optimization process.

Subsequently, we re-purpose the pre-trained speech encoder to generate transcriptions that embed semantic tokens indicative of event tags, optimizing with the CTC loss function. This strategy diverges from previous methods that relied on special characters for tagging within transcriptions using CTC loss. Our approach, which introduces dummy tags during the ASR's pre-training, obviates the need for alterations to the tokenizer or output layer before fine-tuning.

We detail a fine-tuning process for a pre-trained speech encoder that empowers it to recognize utterance intent, demarcate the beginning and end of entities, and carry out transcription, using pairs of speech files and tagged transcriptions. An example of a transcribed speech input is:

\begin{lstlisting}[language=json,firstnumber=1]
@CALENDER_SET@ put !EVENT_NAME! meeting !END! with !PERSON! paul !END! for !DATE! tomorrow !END! !TIME!  ten am !END!
\end{lstlisting}

Here, \emph{CALENDAR\_SET} signifies the intent of the utterance, while the entity tokens \emph{PERSON}, \emph{DATE}, and \emph{TIME} terminate with a shared \emph{END} token. These semantic tokens are mapped to the dummy vocabulary tokens, which are subsequently learned during the fine-tuning stage for the designated task. Although the tokens for utterance level semantics can be placed at the beginning or the end, our empirical evidence suggests a slight performance improvement when it is generated at the beginning.

In our study, we apply the Conformer end-to-end ASR framework \cite{gulati2020conformer}. We use tokenizer output vocabulary $V$ of size 1024, comprised of 400 un-used tokens $D$ for event labels, 624 transcription tokens $L$ and 1 token representating a \emph{Blank} label. These D tokens are provided to tokenizer as a list of \emph{user\_defined\_symbols} 

\begin{figure*}[htbp]
\centering
\includegraphics[width=160mm,height=50mm]{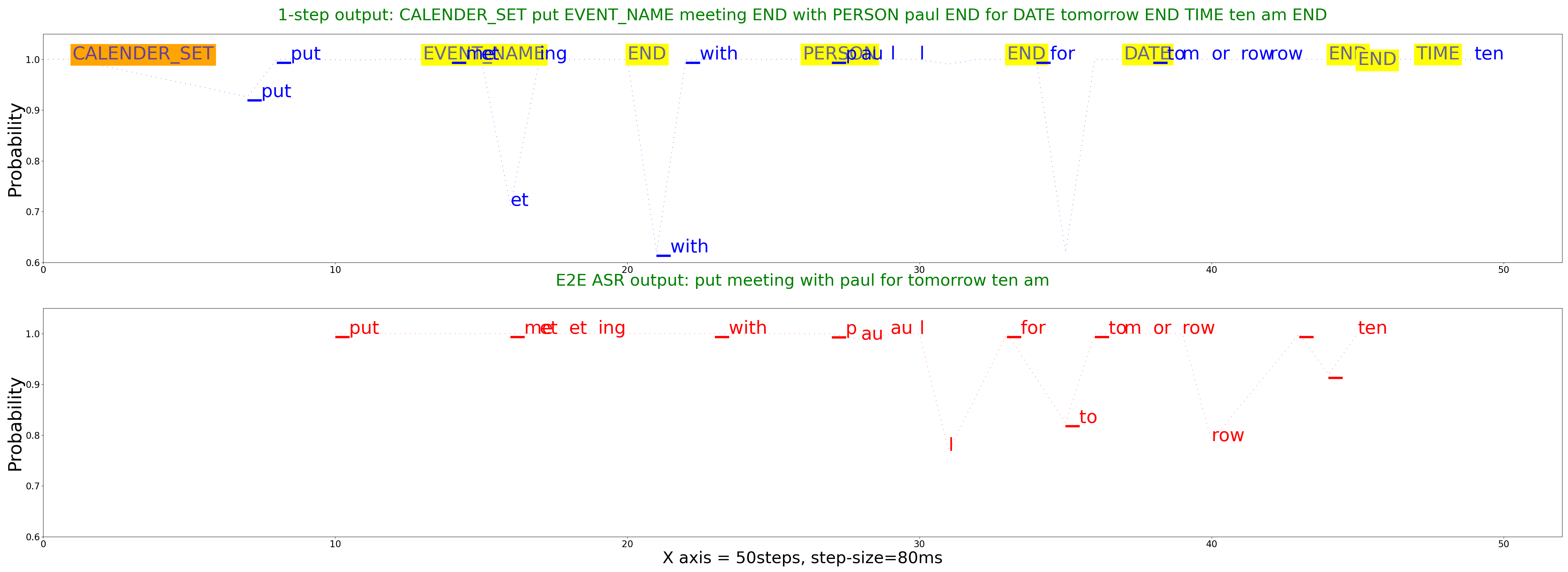}
\caption{Output of softmax probability for token predicted at every time-step. Steps with no visualized token represent "empty" token being predicted. Blue tokens mark our system's output which red tokens represent tokens from pretrained E2E ASR. Our system learns to insert tags by generating less empty tokens and use them to mark event labels instead.}
\label{ctc_fullname}
\end{figure*}

\subsection{From network output to transcriptions with event labels}

 Adopting the computational approach of Connectionist Temporal Classification (CTC), our model is designed to classify sequences of unheard speech with the objective of minimizing the task-specific error metric. This involves correctly predicting output tokens that either correspond to transcription elements or signify an event. Consistent with established CTC methodologies, our network includes a softmax layer that features one additional unit beyond the number of tokens in $V$. This extra unit's activation represents the likelihood of a \emph{blank} or no label. The activations of the first $L$ units are interpreted as the probabilities of each transcription token occurring at specific time steps. The remaining $D$ units are purposed to signal events such as intents, entities or speaker change when activated.

More Formally, for an input sequence x of length $T$(steps in a sample) define a speech DNN encoder with $m$ inputs, $n$ outputs and weight vector $w$ as a continous map $N_w: (R^m)^T \mapsto (R^n)^T$. Let $y = N_w(x)$ be the sequence of network output, and denote by $y_{k}^{t}$ the activation of output unit $k$ at time $t$. $y_{k}^{t}$ is interpreted as the probability of observing label k at time t, thus, defining a distribution over the set $V^{'^T}$ of length T sequences over the alphabet $L^{'} = L \cup \{blank\}$: 

\begin{equation}
    p(\pi|x) = \prod_{t=1}^{T} y_{\pi_t}^t 
\end{equation}

We refer to the elements of the $V^{'^T}$ as paths, and denote them as $\pi$. 

Graves et al. \cite{graves2012connectionist} presuppose in their first equation that outputs of a network at various time steps are independent of each other. Yet, this assumption does not hold in the architecture of modern End-to-End (E2E) speech encoders such as Citrinet, Conformer, and Wav2vec2. These models employ feedback loops that inherently link different temporal outputs, creating a conditional dependency between them. This characteristic interconnection may contribute significantly to the widespread adoption and effectiveness of E2E ASR systems based on CTC methodology.

Many$-$to$-$one map $B$ is defined as $V^{'} \mapsto V^{\leq T}$, where $V^{\leq T}$ refers to set of sequences of length less than or equal to $T$ over the original label vocab V. To obtain read-able transcriptions we apply standard CTC rule of removing all blanks and repeated labels. Finally, $B$ is used to define the conditional probability of an entity $l \in V^{\leq T}$ as sum of probabilities of all the paths corresponding to it:

\begin{equation}
    \sum_{\pi \in B^-1(e)} p (\pi|x)
\end{equation}

The foundational mathematical framework of CTC loss, when combined with the contextualized representations generated by a speech encoder, enables us to refine outputs using previously unused dummy tokens to denote specific event labels in an input stream.

\subsection{Pre-trained checkpoint and fine-tuning details}

We utilize readily available Conformer models \cite{gulati2020conformer} accessible from the NeMo library\footnote{https://tinyurl.com/bdf38tmz}, which have been pre-trained on a corpus comprising 7,000 hours of transcribed public domain audio. These models employ a SentencePiece tokenizer \cite{kudo2018sentencepiece} configured with a vocabulary size of 1024 tokens, denoted as $L$. For the integration of dummy event labels, we conduct an additional phase of fine-tuning focused on transcription tasks. This fine-tuning leverages a dataset amalgamating Fisher \cite{cieri2004fisher}, GigaSpeech \cite{chen2021gigaspeech} and SwitchBoard \cite{godfrey1992switchboard} corpora, which were also part of the initial training set utilized by the pre-trained model checkpoints. In our experiments $L$ event tokens are learnt directly on the event annotated fine-tuning data. 
We use a batch-size of 16 with an initial learning rate of .001. We use a  weight decay of .001 and update the model with 8 accumulated batches with adam back-propagation algorithm. We use 3 A100 GPUs for pre-training and 1 A100 GPU for fine-tuning purposes.The final revision will include more details about the experiment configuration and models.

\section{Dataset}

We evaluate our method on the standardized \emph{SLUE-voxpopuli} \cite{wang2021voxpopuli} and \emph{SLURP} corpus \cite{bastianelli2020slurp}.

\subsection{SLUE}

The SLUE-voxpopuli corpus represents a substantial multilingual collection of spontaneous speech recordings from the European Parliament proceedings. Named Entity (NE) annotations for a 15-hour segment of its training subset and the entire standard development set were recently made available by Shon et al. \cite{shon2022slue}. While the annotations for the test set are not publicly accessible, we are able to evaluate our models on this set by adhering to the submission guidelines provided by the SLUE project\footnote{For submission guidelines, visit: https://asappresearch.github.io/slue-toolkit/}. Details pertaining to the distribution of these data subsets are outlined in Table \ref{tab:SLUE-data}. For the scope of this study, we focus on using combined entity labels, encompassing eight distinct types, rather than employing the more detailed intent labels.

\begin{table}[htbp]
\centering
\scalebox{0.75}{
\begin{tabular}{|cc|cc|}
\hline
\multicolumn{2}{|c|}{\textbf{Label}}                                                                                                             & \multicolumn{2}{c|}{\textbf{Count of phrases}} \\ \hline
\textbf{Raw}                                                                                                      & \textbf{Combined}            & \textbf{Fine-tune}    & \textbf{Dev}           \\ \hline
GPE, LOC,                                                                                                         & PLACE                        & 2012                  & 642                    \\ \hline
\begin{tabular}[c]{@{}c@{}}CARDINAL, \\ ORDINAL\\ QUANTITY, \\ MONEY, \\ PERCENT\end{tabular}                     & QUANT                        & 923                   & 327                    \\ \hline
ORG                                                                                                               & ORG                 & 864                   & 259                    \\ \hline
DATE, TIME                                                                                                        & WHEN                         & 762                   & 260                    \\ \hline
NORP                                                                                                              & NORP                         & 647                   & 220                    \\ \hline
PERSON                                                                                                            & PERSON                       & 272                   & 51                     \\ \hline
LAW                                                                                                               & LAW                          & 250                   & 60                     \\ \hline
\multicolumn{1}{|l}{\begin{tabular}[c]{@{}l@{}}FAC, EVENT, \\ WORK\_OF\_ART,\\ PRODUCT, \\ LANGUAGE\end{tabular}} & \multicolumn{1}{l|}{DISCARD} & \multicolumn{1}{l}{}  & \multicolumn{1}{l|}{}  \\ \hline
\multicolumn{2}{|c|}{TOTAL Entities}                                                                                                             & 5820                  & 1862                   \\
\multicolumn{2}{|c|}{Total Duration}                                                                                                             & 5k (15hrs)            & 1.7k (5hr)             \\ \hline
\end{tabular}
}
\caption{Data Statistics of SLUE-voxpopuli corpus with Named Entity annotations}
\label{tab:SLUE-data}
\end{table}

The evaluation methodology adopted by the SLUE framework entails the transformation of entity occurrences within an utterance into a structured format that includes the entity's \emph{Type}, its \emph{Phrase}, and \emph{Frequency} for both the reference text and the system's output. The selection of the phrase is determined by identifying the beginning and end markers of an entity. In this context, each entity type is signified by a unique beginning token, whereas the end token is standardized across all entity types for the purpose of evaluation. The scores are then computed as follows:

\begin{itemize}
    \item $Recall = Total-correct / Total-reference$
    \item $Precision = Total-correct / Total-system$
    \item $F1score = 2 * (Precision / Recall) / (Precision+Recall)$
\end{itemize}

This evaluation metric does not report the transcription accuracy. Therefore, we report standard WER (after removing semantic event labels) for our experiments.

\subsection{SLURP Corpus}

The Spoken Language Understanding Resource Package (SLURP) dataset comprises a rich collection of 72,000 audio clips, summing up to 58 hours of speech. These recordings simulate single-turn interactions with a virtual home assistant and are annotated across three semantic dimensions: Scenario, Action, and Entities. The dataset is diverse, featuring 56 types of entities, 46 distinct actions, and 18 different scenarios. For automation purposes, action and scenario labels are typically amalgamated to denote the user's intent, resulting in 93 unique intent labels.

The fine-tuning and assessment of the speech encoder leverage the designated official splits for training, validation, and testing from SLURP. We present our findings using the SLURP-specific evaluation metrics, accessible via their official GitHub repository. Unlike the standard SLUE evaluation framework, our entity recognition accuracy is gauged using the provided human-annotated entity boundaries within the audio files. We evaluate system performance by comparing the tuple (Type, Phrase, Frequency) against the reference for each time segment. The calculation of recall, precision, and F-1 score follows the methodology outlined in the preceding section, focusing on the correct identification of entity tags and their corresponding phrases. However, it is important to note that these metrics do not consider the temporal accuracy of the entity tags' start and end points, which is crucial for real-time applications like speech redaction as discussed in Gouvea et al. \cite{gouvea2023trustera}.


\section{Experiments \& Results}
\subsection{SLUE-Voxpopuli}

In the two-step methodology, a text tagger, when applied to human-transcribed speech, achieves the highest accuracy for tagging entities. Past research by Pasad et al. \cite{pasad2021use} indicates that when the Wav2Vec2 (W2V2) model is fine-tuned for both transcription and tagging tasks, there is a noticeable decline in performance. Although Pasad et al. provide results using the W2V2 model, we conducted our experiments using Conformer models. We discovered that employing unused vocabulary labels during the fine-tuning process mitigates the performance degradation. Our proposed method not only improves transcription results but also enhances the accuracy of entity tagging. Our findings also reveal that assigning separate tags for the beginning of an entity is less effective compared to other methods. It is observed that encoders fine-tuned with Connectionist Temporal Classification (CTC) outperform those using the sequence-to-sequence RNNT optimization loss \cite{ghodsi2020rnn}. The baseline results for these observations are documented in the work by Shon et al. \cite{shon2022slue}.

\begin{table}[h!]
\centering
\scalebox{0.75}{
\begin{tabular}{|cc|cc|}
\hline
\multirow{2}{*}{\textbf{Speech Model}} & \multirow{2}{*}{\textbf{Text Model}} & \multirow{2}{*}{\textbf{WER}} & NER-F1        \\
                                       &                                      &                               & \textbf{Comb} \\ \hline
Text NER                               &                                      &                               &               \\
N/A                                    & DeBERTa-B                            & N/A                           & 86.0          \\ \hline
\multicolumn{2}{|l|}{\textbf{2-step Pipeline}}                     &                               &               \\
W2V2-B-LS960  (165m)                   & DeBERTa-B                            & 18.4                          & 78.6          \\
Conformer-CTC  (115m)                      & TPT                                  & 9.8                           &    80.3           \\ 
Conformer-RNNT  (115m)                      & TPT                                  & 9.2                           &    82.1         \\ \hline
\multicolumn{2}{|l|}{\textbf{1-step E2E}}                          &                               &               \\
\multicolumn{2}{|l|}{\uline{Baseline}}                          &                               &               \\

W2V2-B-LS960                           & N/A                                  & 18.4                          & 49.0          \\
Conformer $-$ CTC                              &                                      & 11.0                          & 65.2          \\ 
Conformer $-$ RNNT                              &                                      &         12.1                  &   67.2       \\ \hdashline
\multicolumn{2}{|l|}{\uline{Ours - 1SPU}}                          &                               &               \\

Conformer $-$ CTC                              &        \multirow{2}{*}{N/A}                                & {\bf 9.0}                           & {\bf 74.4}          \\
Conformer $-$ RNNT          &                                      &       10.5                     &       71.1    \\ \hline
\end{tabular}
}
\caption{Results for SLUE benchmark.}
\end{table}

\subsection{SLURP}

\begin{table}[h!]
\centering
\scalebox{0.67}{

\begin{tabular}{|c|ccc|}
\hline
Speech Model & WER & Intent-Acc & Slurp-F1        \\ \hline

\uline{Previous}   & & &\\ 
\textbf{E2E Encoder-Decoder}                          &                                        &    &\\
W2V2-B-LS960                                             &                22.3        &  83.0  &  65.2      \\ 
Conformer-RNNT                                              &           14.2                &  90.1  & 77.22      \\ \hdashline
\textbf{E2E Encoder only}                          &                                       &    & \\
W2V2-B-LS960                                           &                19.2        &  82.0  &  59.2      \\ 
Conformer-CTC                                                   &            15.6               &  83.1  & 66      \\ \hline
\textbf{Proposed 1SPU}                          &                             &          &     \\

Conformer-CTC           &             14.5                &   85.5    &  69.3  \\ \hline

\end{tabular}
}
\caption{Results for SLURP corpus.}
\label{tab:slurp-results}
\end{table}

Table \ref{tab:slurp-results} presents the outcomes for our one-step model and benchmarks it against other E2E approaches. Stutructed prediction Model from NVIDIA\footnote{https://tinyurl.com/nemo-rnnt-structured} (no citation available) , employing an encoder-decoder framework that produces information following a predetermined template, achieves the best performance.  In comparison, our encoder-only model, which leverages CTC loss combined with greedy decoding, outperforms prior models with a similar encoder-only structure. It is important to note, however, that despite the superior performance of the encoder-decoder approach, it yields a structured output, which is not conducive to streamed inference and thus restricts its application to offline processing.


\section{Observations}

Figure \ref{ctc_fullname} illustrates the token output from both a two-step pipeline system and a single-step entity recognition system at each time-step. Tokens from the end-to-end automatic speech recognition (E2E ASR) are marked in red, whereas tokens pertinent to entities identified by the single-step system are highlighted in blue. 

In addition to these experiments, we also experiment with propriety human-human dual channel conversations between a caller and an agent. Assuming negligible overlap in the audio from both parties, we combine the two channels to create a single-channel audio track. To denote transitions between speakers, we introduce a unique label for "Speaker-change." A video demonstration, accessible via the provided link\footnote{https://www.youtube.com/watch?v=10yte4YUqW0}, showcases a sample conversation displayed in a format akin to Figure \ref{ctc_fullname}. From this, we notice an improvement in both the accuracy of entity prediction and overall transcription, which we attribute to the model's capacity to predict semantic events alongside the speech content.

\section{Conclusions}

In our study, we demonstrate that simultaneous high-quality event tagging (encompassing intent, entities, and speaker changes) and transcription is feasible with our proposed method. This technique is adaptable across a broad spectrum of events, making it suitable for extensive applications in speech understanding and transcription tasks.

We are confident that this methodology can be expanded to accommodate additional languages and various domains due to its multi-tasking nature, facilitated by a single, unified loss function. This makes it an ideal candidate for comprehensive, one-step, automated conversational systems. Currently, we are developing a multi-lingual version of this system, which has the capability to perform language translation while concurrently tagging events. Additionally, we aim to refine end-to-end automatic speech recognition (E2E ASR) systems to respond to event recognition prompts based on user-specified natural language queries.

\bibliography{acl_latex}

\end{document}